\documentclass[doublecol]{epl2} 
\usepackage{graphicx}
\usepackage[activeacute,english]{babel}
\begin{document}

\title{The ``footprints'' of irreversibility}

\author{A. Gomez-Marin\inst{1} \and J.M.R. Parrondo\inst{2} \and C. Van den Broeck\inst{3}}
\shortauthor{A. Gomez-Marin \etal}
\institute{ \inst{1} Facultat de F\'isica, Universitat de Barcelona, Diagonal
647, Barcelona, Spain \\ \inst{2} Departamento de F\'{\i}sica At\'{o}mica, Molecular y
Nuclear and {\em GISC}, Universidad Complutense, 28040 Madrid, Spain \\
\inst{3} Hasselt University, B-3590 Diepenbeek, Belgium}

\abstract{ We reformulate the result for the entropy production
given in Phys. Rev. Lett. {\bf 98}, 080602 (2007) in terms of the
relative entropy of microscopic trajectories. By a combination with
the Crook's theorem, we identify the path variables that are
sufficient to fully identify irreversibility. We show that work
saturates the relative entropy, and derive the entropy production
for stochastic descriptions. }



\maketitle

Recent results, known as fluctuation
\cite{fluctuation,fluctuation2,fluctuation3,fluctuation4,fluctuation5}
or work
\cite{work,worka,workb,workc,work2,work2a,work3,work4,work5,work6}
theorems, point to the existence of exact equalities that rule the
fluctuating amounts of work or entropy produced during far from
equilibrium processes. For example, the Jarzynski equality states
that $\langle \exp(-\beta W) \rangle = \exp (-\beta \Delta F)$,
where $W$ is the work needed to bring a system, in contact with a
heat bath at temperature $T$ ($\beta^{-1}\equiv k_B T$), from one
initial state prepared in equilibrium to another one. $ \Delta F$ is
the difference in free energy  of these states (see  \cite{work2}
for a more precise discussion). By the application of Jensen's
inequality, one finds $\langle W \rangle \ge \Delta F$. Since
$(\langle W \rangle - \Delta F)/T$  is the entropy increase in the
entire construction, system plus heat bath, this result is in
agreement with the second law. While such a result is certainly
intriguing and of specific interest for the study of small systems,
where the distribution of work is relevant and measurable, it
provides no extra information on the actual value of the average
work or entropy increase, which is the central quantity in the
second law. Recently however, the microscopically exact value of
these quantities has been obtained in a set-up similar to that of
the work theorem \cite{diss}. The purpose of this  letter is to
investigate some consequences of this result, with special emphasis
on the case when the dynamics of the system can be described in
terms of a reduced set of variables. To make this connection, we
will  rewrite the main result from   \cite{diss}
 in an alternative form, as an integral over
paths. In combination with a microscopic version of Crooks' theorem,
this result identifies the ``footprints'' of irreversibility, namely
the path variables whose statistics are sufficient to reproduce the
exact total entropy production. This prescription  is in agreement
with the expressions for entropy production proposed in the
literature for stochastic models.

We consider a system described by the Hamiltonian
$H(\Gamma,\lambda)$ with $\Gamma=(\{q\},\{p\})$  a point in phase
space, representing all position and momentum variables.  $\lambda$
is an external control parameter, for example the volume or an
external field. The system is perturbed away from its initial
canonical equilibrium at temperature $T$ by changing this control
parameter according to a specific schedule, from an initial to a
final value. For simplicity, we will assume  that  during this time
the system is disconnected from the outside world, except for the
action of changing $\lambda$. This assumption makes the derivation
and discussion simpler, even though the result can be shown to have
a much wider range of validity \cite{vdb}. We also consider the
time-reversed schedule, in which the system starts in canonical
equilibrium at the same temperature $T$, but at the final value of
the control parameter, and  the time-reversed perturbation in
$\lambda$ is applied. We will use the superscript ``tilde'' to refer
to corresponding time-reversed
 quantities (including, by convention, the change of sign for momentum variables).

The quantity of interest is the amount of work $W$ performed during
the forward process. Since the  system is isolated, $W$ is equal to the
energy difference of the system between final and initial state.
While the final state is the deterministic outcome, prescribed by
Hamiltonian dynamics, of the initial condition, the latter is a
random variable in view of the canonical sampling. Therefore $W$ is
also a random variable.  In the following, it will be useful to
regard the work $W$ as a functional of the specific microscopic
trajectory followed by the system. As mentioned above, such a
trajectory is completely specified by the initial condition, but
also by the micro-state $\Gamma$ of the system at any intermediate
time $t$. In \cite{diss}, the following explicit expression was
derived for the corresponding work $W{(\Gamma;t)}$:
\begin{eqnarray} \label{W}
W{(\Gamma;t)} -\Delta F = k_BT \; \ln
\frac{\rho(\Gamma;t)}{\tilde{\rho}(\tilde\Gamma;t)}.
\end{eqnarray}

Here $\rho(\Gamma;t)$ and
$\tilde{\rho}(\tilde\Gamma;t)$ are  the phase space densities at the
same (forward) time $t$  in forward and backward experiment. If the
system is reconnected after the perturbation to a (ideal) heat bath
at temperature $T$, the dissipated work
$W_{\rm{dis}}=W{(\Gamma;t)} -\Delta F$ will be evacuated to the heat
bath, resulting in a  total entropy production equal to
$W_{\rm{dis}}/T$. The above formula is thus the microscopic analogue
of the path-dependent entropy production proposed in various
stochastic models
\cite{crooks,crooksthesis,sekimoto,kurchan,qian,fluctuation5,maes,wu,seifert2005,gaspard2007}.
Our emphasis here however is on the average dissipated work or
average entropy production. Starting from the same Eq.~(\ref{W}), we
derive for this average two different expressions, the combination
of which will lead to a general prescription identifying the
``footprints'' of irreversibility.

First, we derive from Eq.~(\ref {W}) a symmetry  relation for the
probability distribution $P(W)$ of the work as follows:
\begin{eqnarray} \label{PW}
P(W)&=&\langle \delta(W-W{(\Gamma;t)})\rangle\nonumber\\
&=&
\int \rm{d\Gamma}
 \rho(\Gamma;t) \delta(W-W(\Gamma;t))\nonumber\\
&=&
\int \rm{d\Gamma}
 e^{\beta(W(\Gamma;t)-\Delta
F)}\tilde{\rho}(\tilde\Gamma;t)\delta(W-W(\Gamma;t))\nonumber\\
&=&
e^{\beta(W-\Delta
F)}\int \rm{d}\tilde{\Gamma}
 \tilde{\rho}(\tilde\Gamma;t)\delta(W+\tilde W(\tilde{\Gamma};t))
 \nonumber\\
&=& e^{\beta(W-\Delta F)}\;\tilde{P}(-W),
\end{eqnarray}
since the work in the backward processes verifies $\tilde
W(\tilde\Gamma;t)=-W(\Gamma;t)$. This microscopic Crooks' relation
was obtained in the context of Markovian stochastic dynamics by
Crooks \cite{crooks}, and later extended to Hamiltonian dynamics in
\cite{work5}. The above result is usually viewed as an interesting
relation for the probability distribution of the work. It however
also provides a revealing expression for the average work. By
solving Eq.~(\ref{PW}) for $W$ and averaging over $P(W)$, one finds:
\begin{eqnarray} \label{Wcrooks}
\langle W \rangle -\Delta F & = & k_BT\int \rm{d}W\;P(W) \ln
\frac{P(W)}{\tilde{P}(-W)}\nonumber \\
& = & k_B T\; D({P}(W)||{\tilde{P}(-W)}).
\end{eqnarray}
Here, we introduced the relative entropy, or
Kullback-Leibler distance,   between two probability distributions  $p(x)$ and $q(x)$ \cite{cover}:
\begin{equation}
D(p||q)=\int \rm{d}x \; p(x) \ln \frac{p(x)}{q(x)}.
\end{equation}
The relative entropy and its powerful properties will play a central role in the sequel.
At first sight, it may appear superfluous  to express the average $\langle W \rangle$, which is obviously just an integral of $P(W)$, in terms of a more complicated expression involving the second probability distribution $\tilde{P}$ for the reverse experiment. But the following two important properties of the relative entropy \cite{cover}  reveal an additional benefit. Firstly, a relative entropy is non-negative. Eq.~(\ref{Wcrooks}) thus implies that the dissipated work $\langle W \rangle -\Delta F$ is a positive  quantity,
in agreement with the second law.  Secondly, the relative entropy expresses the difficulty for distinguishing samplings from two distributions. The dissipated work is thus equal to the difficulty to distinguish the arrow of time from the statistics of the work involved in forward versus backward experiment.
 The main interest of Eq.~(\ref{Wcrooks})  however comes from its comparison with an expression for $\langle W \rangle$ in terms of the micro-dynamics, which we proceed to derive below.

By performing the straightforward average in Eq. (\ref{W}), we find \cite{diss}:
\begin{eqnarray} \label{maineq00}
\langle W \rangle -\Delta F = k_BT\int \rm{d}\Gamma \rho(\Gamma;t) \ln
\frac{\rho(\Gamma;t)}{\tilde{\rho}(\tilde\Gamma;t)}=k_B T D(\rho
||{\tilde{\rho}}).
\end{eqnarray}
In comparison with Eq.~(\ref{Wcrooks}), the above result fully  reveals the microscopic nature  of the  dissipation, but it
may appear to be of little practical interest. Indeed, it requires
{\it full statistical information} on {\it all} the microscopic
degrees of freedom of the system (even though only at one particular
time).  This stringent requirement is obviously on par with the
generality of the above result, which is valid however far the
system is perturbed away from equilibrium. The perturbation could
therefore imprint its effect on all the degrees of freedom and
their complete statistical properties would be required to reproduce the
corresponding dissipation.

While Eqs.~(\ref{Wcrooks}) and (\ref{maineq00}) provide two
different exact  expressions for the dissipated work, we note that
the  formulas for entropy production in coarse grained  descriptions
are usually in terms of path integrals, on par with the fact that
the determinism of Hamiltonian dynamics is then replaced by
stochastic dynamics. Eq.~(\ref{Wcrooks}) can be considered to be a
path integral version since the work $W$ will, in a reduced
description, indeed depend on the path followed by the coarse
grained variables during the perturbation. To derive a path integral
version of Eq.~(\ref{maineq00}), we invoke another property of
relative entropy \cite{cover}, known as the chain rule. The relative
entropy between probability distributions $p(x,y)$ and $q(x,y)$ of
two random variables can be written as follows:
\begin{eqnarray}\label{cr}
  D\left( p(x,y)||q(x,y)\right) & = &
 D\left(p(x)||q(x)\right)   \\
  & + &  \int\! \rm{d}x \,p(x)\!\int\! \rm{d}y\,
p(y|x)\ln\frac{p(y|x)}{q(y|x)}. \nonumber
\end{eqnarray}
If the random variables are  related to each other by a one-to-one function $x=f(y)$, their conditional probabilities become infinitely sharp and the
second term in the r.h.s. of Eq.~(\ref{cr}) vanishes.
One then finds:
\begin{equation}\label{rei}
D\left( p(x,y)||q(x,y)\right)=D\left( p(x)||q(x)\right)=D\left(
p(y)||q(y)\right).
\end{equation}
In words, the addition of dependent variables does not modify the relative entropy.
Since Hamiltonian  dynamics generates such one-to-one
relations between the micro-states at different times, one can specify in Eq.~(\ref{maineq00}), without changing
the value of the relative entropy, the micro-state $\Gamma_i$ of the
system at as many additional measurement points in times  $t_i$, $
i=1,...,n$, as one likes:
\begin{eqnarray} \label{maineq001}
\langle W \rangle -\Delta F = k_BT\int \prod_{i=1}^{n}
d\Gamma_i
 \;\rho(\{\Gamma_i;t_i\}) \ln
\frac{\rho(\{\Gamma_i;t_i\})}{\tilde{\rho}(\{\tilde{\Gamma}_i;t_i\})}.
\end{eqnarray}
In the continuum limit covering the entire time interval (with $n
\rightarrow \infty$), one thus converges to the following result in terms of a path integral, see also \cite{jarzynski2006}:
\begin{eqnarray}
\langle W \rangle -\Delta F & = & k_B T\int {\cal D}({\mbox{path}}) {\cal
P}(\mbox{path})\ln \frac{ {\cal P}(\mbox{path})}{\widetilde{{\cal
P}} (\widetilde{\mbox{path}})}\nonumber\\
& = &  k_B  T { D}({{\cal
P}(\mbox{path})}||{\widetilde{{\cal P}}
(\widetilde{\mbox{path}})})\label{maineq}.
\end{eqnarray}
This expression, while containing redundant information from the
point of view of Hamiltonian dynamics, has the important advantage
 that it is also exact and formally identical, as shown below,  when the
paths are expressed no longer in terms of microscopic variables but
in terms of an appropriate set of reduced variables. Furthermore, in the latter case, the path formulation is no longer
redundant since the trajectory captures information about the
eliminated degrees of freedom.
The identification of the minimal
set of variables, for which the elimination is valid, follows from the combination of Eq.~(\ref{maineq}) with
the Crooks'  result Eq.~(\ref{Wcrooks}). One finds:
\begin{equation} \label{dwdp}
D( {\cal P} (\mbox{path}) || \widetilde{{ \cal
P}}(\widetilde{\mbox{path}})) = D(P(W)||\widetilde P(-W)).
\end{equation}
 This is a surprising relationship:   from the chain rule for the relative entropy,
Eq.~(\ref{cr}), one would expect that the relative entropy for the
paths, which contains full information on all the microscopic variables, would be bigger than that contained in the work, which is a single scalar path-dependent variable. However, both relative entropies are exactly the same. The combination of Eqs.~(\ref{maineq}) and ~(\ref{dwdp})   allows us now to formulate the following main conclusions. First,  it is impossible via relative entropy to overestimate the dissipation. Second, the exact dissipation  is revealed by any set of  variables that contains the statistical information about the work. The fact that the dissipation is underestimated if we do not have this information is also of practical interest, but will be the object of a separate paper \cite{gomeznew}.

One set of variables that captures the information on the work is
now easy to identify: the information is obviously  contained in the
dynamic variables that are interacting with the (external)
work-performing device. More precisely, the work performed along a
trajectory $\Gamma(t)$, $t\in [0,\tau]$, is given  by:
\begin{equation} W=\int_0^{\tau}dt \frac{\partial H(\Gamma(t),\lambda(t))}{\partial
\lambda(t)}\dot\lambda(t)
\end{equation}
and can be exactly calculated from the path followed by the
variables coupled to $\lambda$.
Then it is enough to know the
(statistical) behavior of these variables to reproduce the
statistics of the work, and hence the average dissipation.
Trajectory information of these and only these variables, along the
whole (both forward and backward) process, is enough to account
for the total average dissipation. In particular, if a stochastic
model provides the exact description of a system in its interaction
with an external device, one needs only the path information of
these variables. Eq.~(\ref{maineq}) is thus valid for a ``correct''
stochastic model with the path determined in terms of the
corresponding stochastic variables. As a corollary, we note that
bath variables which are replaced (in some ideal limit) by a
stochastic perturbation, will not appear in the ``path'', which is
in terms of the variables of the stochastic system only.

Let us mention another surprising consequence of the above equality (\ref{dwdp}). By applying the chain rule, Eq.~(\ref{cr}), one finds:
\begin{equation}\label{d0} D( {\cal P} (\mbox{path}|W) || \widetilde{{ \cal
P}}(\widetilde{\mbox{path}}|-W))=0,
\end{equation}
and hence:
\begin{equation}\label{d02}  {\cal P} (\mbox{path}|W) = \widetilde{{ \cal
P}}(\widetilde{\mbox{path}}|-W),
\end{equation}
for all $W$.  Eq.~(\ref{d0}) means  that, by selecting trajectories
corresponding to a given value of work, $W$ and $-W$,  in the
forward and backward process respectively,  it is not possible to
detect the arrow of time in them. According to Eq.~(\ref{d02}), the
sub-ensembles of these trajectories are in fact statistically
indistinguishable! As an example, the snapshots of the positions of
particles, during the expansion and compression of a gas, will be
statistically identical, when the corresponding amounts of work are
each other's opposite. For the folding or unfolding of an RNA
molecule \cite{ritort}, the trajectories are statistically
indistinguishable, again if the amounts of works are opposite. We
however also note that, according to the Crooks' relation
Eq.~(\ref{PW}),  the probabilities for  such forward and backward
trajectories will be very different if the experiment is performed
in an irreversible way.  If, e.g., the forward set  corresponds to
typical realizations, the same set of trajectories will be
atypical for the backward experiment \cite{jarzynski2006} (except
if the overall process is reversible).

We conclude with a brief discussion about  the range of
applicability of the above result. Recall that, in its derivation,
it is assumed that the system starts in canonical equilibrium in
both forward and backward scenario, and is disconnected from the
heat bath in the intermediate time.  However, both the microscopic
expression for dissipation given in Eq. (\ref{maineq00}) and the
Crooks' equality can be derived for other equilibrium initial
conditions \cite{vdb} so that Eq.~(\ref{maineq}) remains valid for
these other  ``transient'' nonequilibrium scenarios linking
equilibrium states. Furthermore, the formula can also be applied to
nonequilibrium steady states according to the following argument.
Imagine that the  perturbation induces, after an initial transient,
a steady state in a sub-part of the system. In the limit that the
other degrees of freedom for the remainder of the system have an
infinitely fast relaxation to (local) equilibrium, these will not
contribute in the formula and both the time-irreversibility and
dissipation will be completely captured by the steady state
variables. This hand-waving argument explains why the  formula
(\ref{maineq}) is also known to reproduce the correct entropy
production in nonequilibrium steady state models  \cite{luo,blythe},
where ideal heat, work, and/or particle sources are responsible for
the generation of the steady state.

\acknowledgments
We acknowledge financial help from the Ministerio de Educacion
y Ciencia (Spain) under Grants FPU-AP-2004-0770 (A. G-M.) and
MOSAICO (JMRP), and from the FWO Vlaanderen.

\end{document}